\documentclass[conference,letterpaper]{IEEEtran}

\usepackage{graphicx}
\usepackage[utf8]{inputenc} 
\usepackage[T1]{fontenc}
\usepackage{url}
\usepackage{ifthen}
\usepackage{color}

\usepackage[cmex10]{amsmath} 
\usepackage{amssymb}
\usepackage{amsthm}
\usepackage{algorithm}
\usepackage{algorithmic}
\usepackage{soul}

\def \({\left(}
\def \){\right)}
\def \[{\left[}
\def \]{\right]}
\newcommand{\tbf}[1]{{\textbf{#1}}}

\newcommand{\defeq}{\vcentcolon=}

\newcommand{\bA}{{\textbf {A}}}

\newcommand{\bx}{{\textbf {x}}}
\newcommand{\bg}{{\textbf {g}}}

\newcommand{\by}{{\textbf {y}}}

\newcommand{\bS}{{\textbf {S}}}

\newcommand{\be}{\begin{equation}}
\newcommand{\ee}{\end{equation}}
\newcommand{\bea}{\begin{eqnarray}}
\newcommand{\eea}{\end{eqnarray}}

\makeatletter
\renewcommand*{\defeq}{\mathrel{\rlap{%
                     \raisebox{0.3ex}{$\m@th\cdot$}}%
                     \raisebox{-0.3ex}{$\m@th\cdot$}}%
                     =}
\makeatother

\begin{document}

\title{Sparse superposition codes under VAMP decoding with generic  rotational invariant coding matrices}
\author{\IEEEauthorblockN{TianQi Hou$^*$, YuHao Liu$^{\dagger}$, Teng Fu$^{\dagger}$ and Jean Barbier$^{ \diamond }$\\
$*$ Theory Lab, Central Research Institute, 2012 Labs, Huawei Technologies Co., Ltd. \\
$\dagger$ Department of Mathematical Sciences, Tsinghua University, Beijing, China    \\
$ \diamond $  International Center for Theoretical Physics, Trieste, Italy\\
} Emails: thou@connect.ust.hk, \{yh-liu21, fut21\}@mails.tsinghua.edu.cn, jbarbier@ictp.it}
\maketitle
\IEEEpeerreviewmaketitle

\begin{abstract}
Sparse superposition codes were originally proposed as a capacity-achieving communication scheme over the gaussian channel, whose coding matrices were made of i.i.d. gaussian entries \cite{barron2010toward}. We extend this coding scheme to more generic ensembles of rotational invariant coding matrices with arbitrary spectrum, which include the gaussian ensemble as a special case. We further introduce and analyse a decoder based on vector approximate message-passing (VAMP) \cite{rangan2019vector}. Our main findings, based on both a standard replica symmetric potential theory and state evolution analysis, are the superiority of certain structured ensembles of coding matrices (such as partial row-orthogonal) when compared to i.i.d. matrices, as well as a spectrum-independent upper bound on VAMP's threshold. Most importantly, we derive a simple ``spectral criterion'' for the scheme to be at the same time capacity-achieving while having the best possible algorithmic threshold, in the ``large section size'' asymptotic limit. Our results therefore provide practical design principles for the coding matrices in this promising communication scheme.    
\end{abstract}

\section{Introduction and setting}
Sparse superposition (SS) codes were introduced for communication over the additive white gaussian noise channel (AWGNC) \cite{barron2010toward} and proven to achieve the capacity using power allocation \cite{rush2017capacity} or spatial coupling under message-passing based decoding \cite{barbier2017approximate,barbier2016proof,biyik2017generalized,rush2021capacity}. But the coding matrices were limited to be constructed from independent gaussian entries. In this paper, we extend the coding matrices to a much broader class of matrices beyond the i.i.d. ones, i.e., to rotational invariant matrix ensembles. We deal with several illustrative coding ensembles, both theoretically and practically, by introducing and analyzing a VAMP-based decoding algorithm \cite{rangan2019vector} (which is similar to OAMP \cite{ma2017orthogonal}). Furthermore, we empirically confirm that a state evolution (SE) recursion  accurately tracks VAMP's performance. By analyzing the fixed points of this SE recursion combined with a replica analysis from statistical mechanics, we precisely quantify computational-to-statistical gaps for several coding matrix ensembles, and demonstrate the superiority of coding matrices whose rows are orthogonal compared to the standard gaussian coding ensemble. Other important contributions come in the form of a simple criterion to select ``good'' coding matrices, i.e., for the coding scheme to be capacity-achieving \emph{and} with the best possible algorithmic threshold, in the large section size limit, but also a spectrum-independent upper bound on VAMP's algorithmic threshold which sets an absolute limit on its performance for any rotational invariant ensemble.               

Let us emphasize that all our results are at the moment non-rigorous. Our main tools are the replica symmetric method \cite{mezard2009information} and the state evolution recursion tracking AMP-like algorithms \cite{bayati2011dynamics,javanmard2013state,bayati2011dynamics}. Concerning the replica method, despite being non-rigorous, a multitude of recent studies  prove its exactness in many similar inference problems \cite{rush2017capacity,barbier2020mutual,reeves2016replica,barbier2016threshold,barbier2019optimal,barbier2018mutual}. This strongly points towards the fact that our replica-based predictions should be exact in a proper asymptotic limit. For the state evolution analysis, it is proven to track VAMP but only for separable denoisers (that would correspond to the trivial $B=1$ case of SS codes). Extending the VAMP state evolution to section-wise priors as needed here requires some work, in the spirit of \cite{berthier2020state} for AMP. Even more care is needed when considering the ``large section size limit'' that we are also going to study; see \cite{rush2017capacity} where this was done for the standard SS codes with gaussian coding matrices under AMP decoding. Nevertheless, we conjecture that all the present results can, and will, be proven in the future. We also empirically confirm our predictions through careful numerics. Therefore our results must be considered as numerically-verified conjectures based on by-now well established techniques from statistical physics and the theory of message passing algorithms.


 In SS codes, the \emph{message} $\bx \!=\! [\bx_1, \dots, \bx_L]$ is a vector made of $L$ sections, each with $B$ entries. Each section $\bx_l$, $l\!\in\!\{1,\ldots,L\}$ possesses a single non-zero component equal to $1$ whose position encodes the symbol to transmit. $B$ is the \emph{section size} (or alphabet size) and we set $N\defeq LB$. We consider random codes generated by a \emph{coding matrix} $\bA\!\in\! \mathbb{R}^{M \times N}$ drawn from a rotational invariant ensemble, i.e., when considering its singular value decomposition $\bA = \textbf{U} \sqrt{\textbf{D}} \textbf{V}^\intercal  $, the orthogonal basis of singular vectors  $\textbf{U}$ and $\textbf{V}$  are sampled uniformly in the orthogonal group $\mathcal{O}(M)$ of $M\times M$ matrices and $\mathcal{O}(N)$, respectively. The diagonal matrix $ \textbf{D}$  contains non-negative singular values $(\textbf{D}_{i})_{i \leq N}$ on its main diagonal, and whose empirical distribution  $N^{-1} \sum_{i \leq N} \delta_{\textbf{D}_{i}} $ weakly converges to a well-defined compactly supported probability density function $\nu(\lambda)$ as $ N, M \to \infty$ (not necessarily proportionally). We denote its aspect ratio $\alpha=M/N $. The cardinality of the code is $B^L$. Hence, the (design) rate is $R=L\log_2(B)/M=\log_2(B)/(\alpha B)$ and thus the code is fully specified by $(M, R, B)$. For a message $\bx$ as before, the \emph{codeword} is $\bA\bx\!\in\! \mathbb{R}^{M}$. We enforce the power constraint { $\|\bA\bx\|_2^2/M=1+o_L(1)$ by tuning the spectrum  $ \nu(\lambda) $ to satisfy $ \int d\lambda \lambda \nu(\lambda)= \alpha B$}. Codewords are transmitted through an AWGNC, i.e., the
received corrupted codeword is $ \by=\bA \bx+  \textbf{z}  $, with i.i.d. $ z_{\mu}  \sim \mathcal{N}\left(0, \sigma^{2}\right), \mu \le M $,  so that the signal-to-noise ratio is $ \text{snr}= \sigma^{-2}$.

\section{The VAMP-based decoder}
The VAMP algorithm we propose, see Algorithm 1 below, aims at computing the minimum mean-square (MMSE) estimator given by the expectation of the Bayesian posterior
\begin{equation*}
   P(\textbf{x} \mid  \textbf{y}, \textbf{A} ) = \frac{1}{\mathcal{Z}(\textbf{y}, \textbf{A})} 
     \exp \Big(-\frac{\text{snr}}{2} \| \textbf{A} \textbf{x} - \textbf{y}\|_2^2\Big) \prod_{l\le L} P_0(\textbf{x}_l), 
\end{equation*}
where $\mathcal{Z}(\textbf{y}, \textbf{A}) $ is a  normalization. But as we will see it is successful in doing so only for certain rates $R$. The hard constraints for the sections of the message are enforced by the prior distribution $P_0(\bx_l)=B^{-1}\sum_{i\in l} \delta_{x_i,1}\prod_{j\in l, j\neq i} \delta_{x_j,0}$, where $\{i\!\in\!l\}$ are the $B$ scalar components indices of the section belonging to the section indexed by $l$.

\begin{algorithm}[t]
\newcommand{\Kit}{K}
\newcommand{\kp}{k\!+\!}
\caption{VAMP-based decoder for SS codes}
\begin{algorithmic}[1]  \label{alg:vamp_slm}
\REQUIRE{Max iteration $K$, coding matrix $\bA$, observed $\by$  }
\STATE{ Initialize  $ \textbf{r}_{1,0} $ and $\gamma_{1,0}  \geq 0 $.             }
\FOR{$k=0,1,\dots,\Kit$ (or until convergence)}
    \STATE{// Denoising}
    \STATE{$ \hat{\textbf{x}} _{1,k} = \textbf{g}_1(\textbf{r}_{1,k},\gamma_{1,k})$,~~
           $\alpha_{1,k} = \left\langle\textbf{g}_1'(\textbf{r}_{1,k},\gamma_{1,k})\right\rangle $}
        \label{line:x1}
    \STATE{$\textbf{r}_{2,k} = (\hat{\textbf{x}}_{1,k} - \alpha_{1,k} \textbf{r}_{1,k})/(1-\alpha_{1,k})$}
        \label{line:r2}
    \STATE{$\gamma_{2,k} = \gamma_{1,k}(1-\alpha_{1,k})/\alpha_{1,k}$}
        \label{line:gam2}
    \STATE{// LMMSE estimation}
    \STATE{$ \hat{\textbf{x}} _{2,k} = \textbf{g}_2(\textbf{r} _{2,k},\gamma_{2,k})$,~~
           $ \alpha_{2,k} =\left\langle\textbf{g}_{2}^{\prime}\left(\mathbf{r}_{2,k}, \gamma_{2, k}\right)\right\rangle    $  }
        \label{line:x2}
    \STATE{$\textbf{r}_{1,\kp1} = (\hat{\textbf{x}}_{2,k} - \alpha_{2,k}\textbf{r}_{2,k})/(1-\alpha_{2,k})$}
        \label{line:r1}
    \STATE{$\gamma_{1,\kp1} = \gamma_{2,k}(1-\alpha_{2,k})/\alpha_{2,k}$}
        \label{line:gam1}
\ENDFOR
\STATE{Return $\hat \bx=\hat{\textbf{x}}_{1,\Kit}$.}
\end{algorithmic}
\end{algorithm}

VAMP was originally derived for generalized linear estimation \cite{rangan2019vector}. In the present generalization to the vectorial setting of SS codes, only the input non-linear steps differ from canonical VAMP: here the so-called denoiser $ \textbf{g}_{1} ( \textbf{r},\gamma) $ acts section-wise instead of component-wise. In full generality, it is defined as  $\textbf{g}_{1} ( \textbf{r},\gamma   )\!\defeq\!\mathbb{E}[\textbf{X} \mid \textbf{R}\!=\!\textbf{r}]$ for the random variable $\textbf{R}\!=\!\textbf{X}\!+\! \sqrt{\gamma}\,{\textbf{Z}}$ with $\textbf{X}\!\sim\!P_0^{\otimes L}$ and ${\textbf{Z}}\!\sim\!{\cal N}(0,  \tbf{I}_N)$. Plugging $P_0$ yields the component-wise expression of the denoiser and its variance:
\begin{align*}
	\begin{cases}
	[ \textbf{g}_{1}(\textbf{r},  \gamma )]_i& = \frac{\exp(r_i/ \gamma)}{\sum_{j\in l_i}\exp(r_j/ \gamma)}, \nonumber\\
	 [ \textbf{g}_{1}'(\textbf{r},  \gamma )]_i  &=\gamma^{-1} [ \textbf{g}_{1}'(\textbf{r},  \gamma )]_{i}( 1- [  \textbf{g}_{1}'(\textbf{r},  \gamma )]_{i} ),
	\end{cases}
\end{align*}
where $[\textbf{g}_{1}^{\prime}(\textbf{r},  \gamma )]_i:=\partial_{{x_i}} {\bg}_{1}({{\boldsymbol{x}}},  \gamma)|_{{{x_i}}=r_i}$, $l_i$ is the section to which belongs the $i^{\text{th}}$ scalar component.   $\textbf{g}_{2} ( \textbf{r},\gamma   )$ can be recognized as the MMSE estimate of a random vector $\textbf{x}$ from the data $\by\sim \mathcal{N}( \mathbf{A} \bx, \text{snr}^{-1} \mathbf{I}_N) $ and prior $ \bx \sim \mathcal{N}(\mathbf{r}, \gamma^{-1} \mathbf{I}_N)$: 
\begin{align*}
\begin{cases}
	  \bg_{2}(\mathbf{r}, \gamma)&=( \text{snr}  \mathbf{A}^{\intercal} \mathbf{A}+\gamma \mathbf{I}_N)^{-1}(\text{snr}\mathbf{A}^{\intercal} \mathbf{y}+\gamma \mathbf{r}),      \\
	  \langle\bg_{2}^{\prime}(\mathbf{r}, \gamma)\rangle &=\gamma N^{-1} \operatorname{Tr}[( \text{snr} \mathbf{A}^{\intercal} \mathbf{A}+\gamma \mathbf{I})^{-1}].
	  \end{cases}
\end{align*}
We will track two error metrics for the VAMP estimator $\hat \bx=(\hat{\bx}_{1}, \ldots, \hat{\bx}_{L})=(\hat{x}_{1}, \ldots, \hat{x}_{N})$, namely the mean-square error (MSE) per section $E_L$ and the section error rate $\mathrm{SER}_L$ ($ \mathbb{I}(\cdot) $ is the indicator):
\begin{equation*}
    E_L\defeq \frac{1}{L}\|\bx-\hat{\bx}\|_2^{2}, \quad \mathrm{SER}_L\defeq \frac{1}{L} \sum_{l\le L} \mathbb{I}\left(\bx_{l} \neq \hat{\bx}_{l}\right).
\end{equation*}

\section{State evolution and replica analyses}
We now present the state evolution (SE) and replica symmetric analyses, valid in the asymptotic  $L\!\to\!\infty$ limit, of the performance of SS codes under MMSE and VAMP decoding, for rotational invariant coding matrices. Both analyses are intimately related, and linked to the estimation problem of a single section $\textbf{S}\!\sim\! P_0$ transmitted through an ``effective gaussian channel''  with noise variance $$\Sigma(E)^2\defeq (B \text{snr} \mathcal{R}(-\text{snr} E))^{-1}.$$ Here $ \mathcal{R}(z)\defeq \mathcal{C}^{-1}(-z)-z^{-1}$ is the R-transform associated to the asymptotic spectral density $\rho$ of $B^{-1} \textbf{A}^\intercal \textbf{A}$, where $\mathcal{C}^{-1}$ is the functional inverse of the Cauchy transform $  \mathcal{C}(z):=  \int\frac{\rho(\lambda)}{\lambda-z}  d \lambda$, see, e.g., \cite{tulino2004random,potters2020first}.

\begin{figure}
    \centering
    \includegraphics[trim={98 40 98 70},clip,scale=0.44]{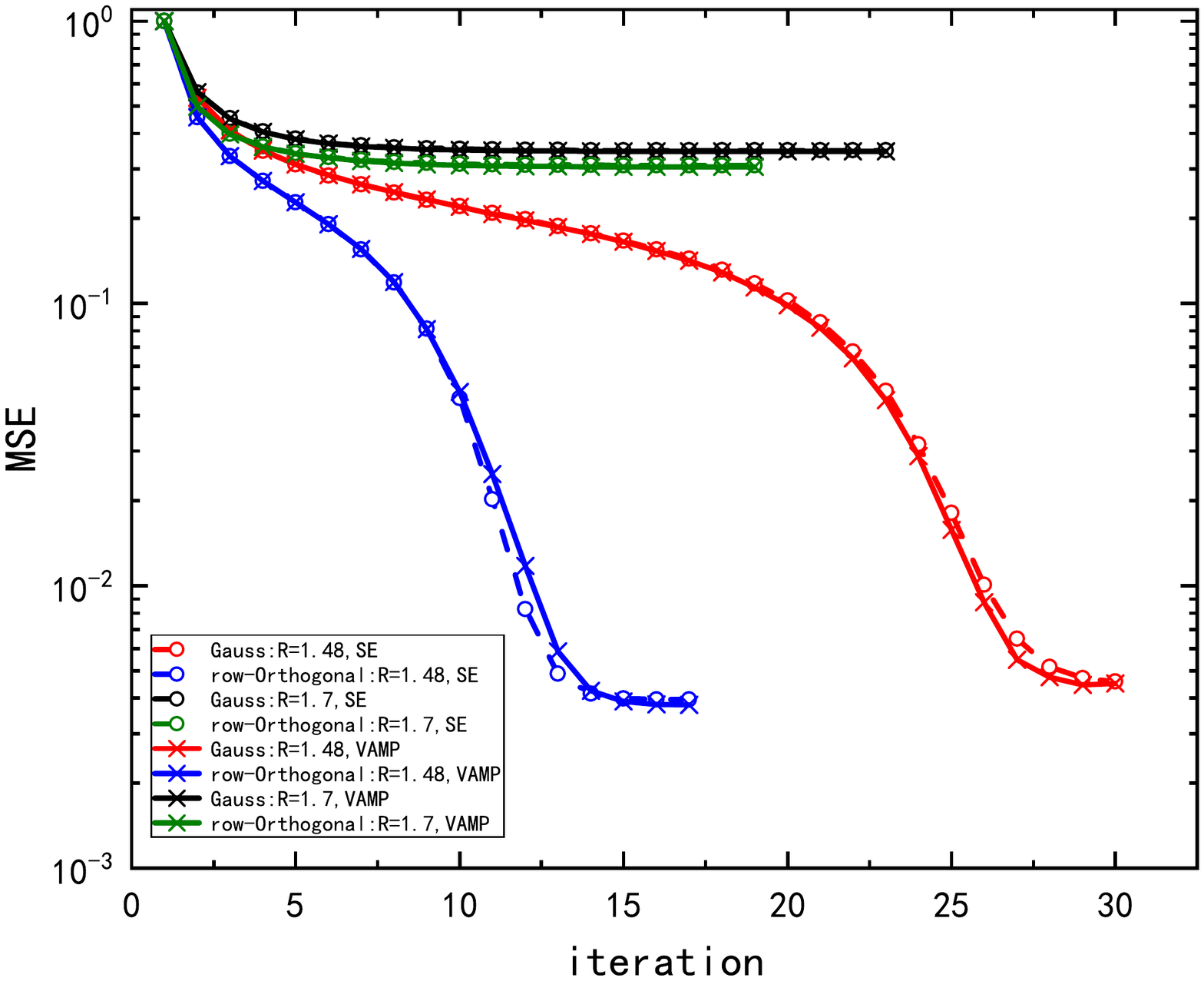}
    \caption{State evolution (dashed dotted curves) tracking the VAMP decoder MSE (solid crossed lines) {{ran}} on single instances of size $L=2^{16}$ for $\text{B}=4$ and $\text{snr}=28$, as a function of the iterations. The SE is computed using Monte Carlo integration with $10^5$ samples. Two types of coding matrices are considered: standard coding  matrices with i.i.d. gaussian entries, and partial row-orthogonal ones, in both cases for rates smaller and larger than their respective algorithmic thresholds $R_{\text{VAMP}}$.}
    \label{fig:SE of VAMP}
\end{figure}

\emph{State evolution} \ \  Let the scalar-valued SE operator 
\begin{align*}
T(E) &\defeq \mathbb{E}_{\textbf{S},\textbf{Z}}\|\textbf{S}\!-\! \mathbb{E}[\textbf{S}\mid \textbf{S}\!+\!\Sigma(E)\textbf{Z}]\|_2^2\\
&=\mathbb{E}_\textbf{Z}[(g^{(1)}(\Sigma(E),\textbf{Z})\!-\!1)^2\!+\!(B\!-\!1)g^{(2)}(\Sigma(E),\textbf{Z})^2]
\end{align*}
where $\textbf{S}\sim P_0, \textbf{Z}\sim \mathcal{N}(0,\tbf{I}_B)$ and we define 
\begin{align*}
	\begin{cases}
	g^{(1)}(\Sigma,\textbf{z}) &\defeq [1\!+\! e^{-\frac{1}{\Sigma^2}}\sum_{j=2}^B e^{\frac{1}{\Sigma}(z_j-z_1)}]^{-1},\\
	g^{(2)}(\Sigma,\textbf{z}) &\defeq [1\!+\!e^{\frac{1}{\Sigma^2}+(z_1-z_2)\frac{1}{\Sigma}}\!+\!\sum_{k=3}^B{e^{(z_k-z_2)\frac{1}{\Sigma}}}]^{-1}.
	\end{cases}
\end{align*}
The SE operator corresponds to the MMSE of $\bS$ when transmitted over the effective gaussian channel. The SE recursion tracking the asymptotic $L\to\infty$ limit $E^{(t)}$ of VAMP's MSE at iteration $t$ is then obtained as a straightforward adaptation of the results of \cite{rangan2019vector} and reads
\begin{align}
E^{(0)} = 1-1/B, \quad E^{(t+1)} = T(E^{(t)}), \quad t\geq 0.	\label{SE}
\end{align}
We refer to \cite{rangan2019vector} for the proof of VAMP's state evolution in regression. Fig. \ref{fig:SE of VAMP} is a numerical demonstration of the validity of our SE recursion for tracking VAMP for SS codes.

\emph{Replica symmetric analysis} \ \ A remarkable property of the SE recursion \eqref{SE} is that its stationary point(s) are in one-to-one correspondence with the critical point(s) of the so-called {\it replica symmetric potential} (or ``free entropy'') $\Phi_{B}(E)$ derived from the replica method \cite{mezard2009information,tulino2013support,barbier2017approximate}: $$\partial_E \Phi_B(E)|_{E_*}=0 \Leftrightarrow T(E_*) = E_*.$$ For the SS codes in the present setting it is:
\begin{align*}
	\begin{cases}
	\Phi_{B}(E) &\defeq  S_{B}(\Sigma(E))\!-\! U_{B}(E),\\
	U_{B}(E) &\defeq   \frac{B}{2}\int_{0}^{ \text{snr} E   } \mathcal{R}(x)dx -\frac{E}{2 \Sigma^2(E)} ,\\
	S_{B}( \Sigma( {E})) &\defeq \mathbb{E}_{\textbf{Z}}\log_B\big(1+\sum_{i=2}^B e_{i}(\textbf{Z},\Sigma(E))\big),
	\end{cases}
\end{align*}
where $e_{i}(\textbf{Z},x)\! \defeq\! \exp((Z_i\!-\!Z_1)/x\!-\!1/x^{2} )$ and i.i.d. $Z_i\!\sim\!\mathcal{N}(0,1)$. The validity of such replica analysis has by-now been proven in many related settings to ours, such as generalized linear regression and compressive sensing but with gaussian i.i.d. matrices \cite{barbier2020mutual,barbier2019optimal,reeves2016replica}. Apart from few recent works  \cite{barbier2018mutual,gerbelot2020asymptotic,fan2022approximate,dudeja2020information,venkataramanan2021estimation}, the rigorous study of linear regression problems with rotationally invariant matrix ensembles is only at its premises. Thus, proving our present conjectures is an interesting avenue left for future work.

We illustrate our results through three coding ensembles: 

$\bullet$ $(i)$ As base case we consider the standard gaussian setting where all entries of  $ \bA$ are i.i.d. gaussian. The asymptotic results turn out to depend only on the asymptotic eigenvalue distribution $\rho(\lambda) $ of $B^{-1} \bA^{T}\bA$, which is given in this case by the Marchenko–Pastur distribution $ \rho(\lambda)=(1-\alpha) \delta(\lambda) +\sqrt{(\lambda-\lambda_{-})(\lambda_{+}-\lambda)}/(2 \pi \lambda)$, where $  \lambda_{\pm} \defeq (1 \pm \sqrt{\alpha})^2 $. Then $ \mathcal{R}(z)=\alpha/(1-z)$ \cite{tulino2004random}. As it should, the analysis does recover in this case the results of \cite{barbier2014replica,barbier2017approximate,rush2017capacity}. 

$\bullet$ $(ii)$ The row-orthogonal ensemble constructed by randomly selecting $M\le N$ rows
from a uniformly sampled $N \times N $ orthogonal matrix. For this ensemble $ \rho(\lambda)=   (1-\alpha) \delta(\lambda) + \alpha \delta (\lambda-1)    $ and  $ \mathcal{R}(z)=(1+z+\sqrt{(1-z)^{2}+4 \alpha z})/(2 z)-z^{-1} $. This will recover similar  results as found in \cite{ma2014turbo}.

$\bullet$ $(iii)$ Finally, a discrete spectrum $ \rho(\lambda)=  (1-\alpha) \delta(\lambda)+\frac{\alpha}{2} \delta(\lambda-\frac{1}{2})+\frac{\alpha}{2} \delta(\lambda-\frac{3}{2}) $. This ensemble is obtained by generating a spectrum with the proper fractions of singular values of $\bA$ in $\{0,\sqrt{1/2}, \sqrt{3/2}\}$ and then multiplying  by uniform orthogonal matrices of proper dimensions. The R-transform is obtained as the solution of a cubic equation solved numerically. This rather artificial case will serve as tractable example of a \emph{non} capacity-achieving ensemble.

\begin{figure}[t]
    \centering
    \includegraphics[trim={104 40 98 70},clip,scale=0.43]{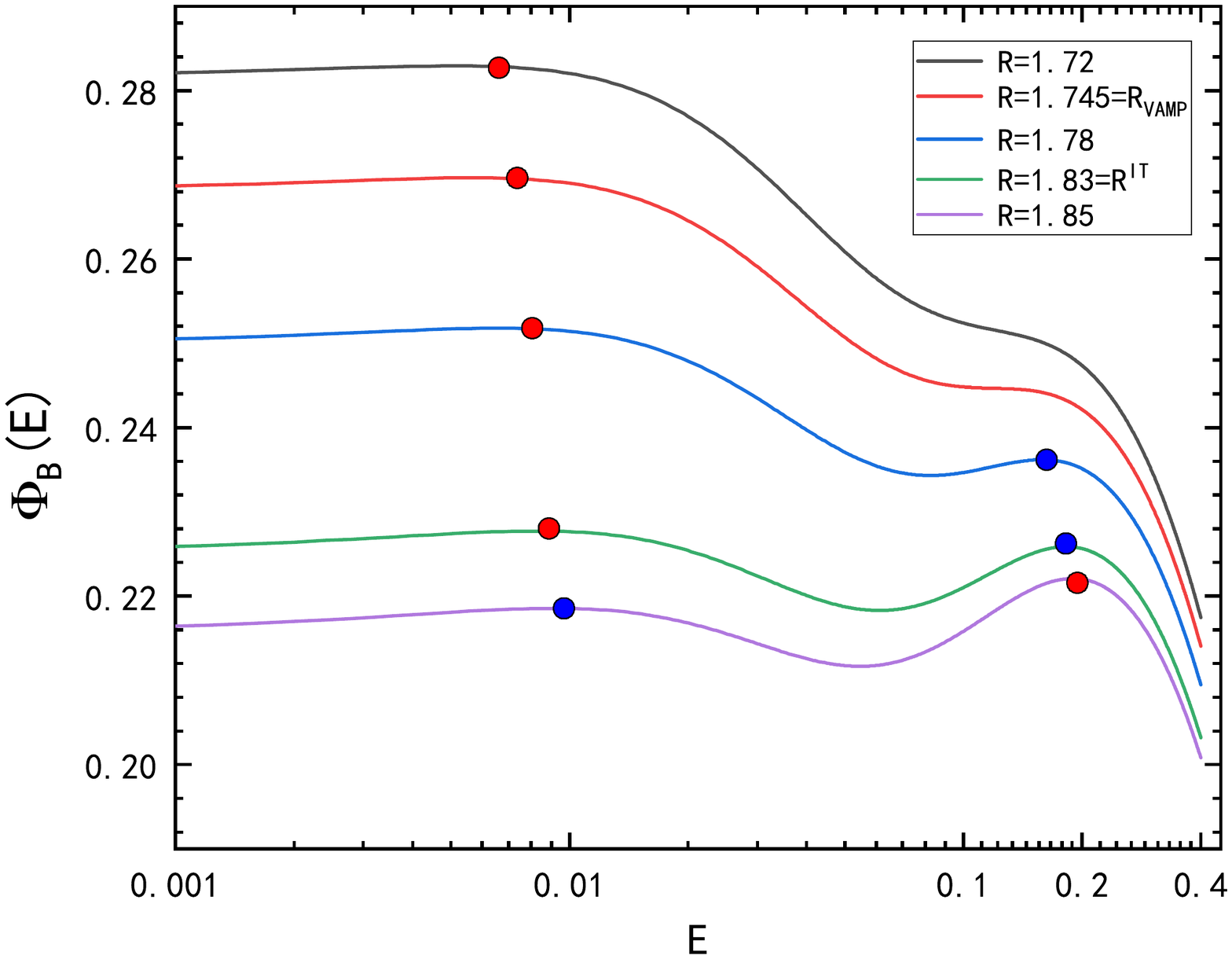}
    \caption{Replica symmetric potential $\Phi_{B}(E)$ with row-orthogonal coding matrix, $\text{snr}=28$ and $B=2$. The red dot is the global maximum while the blue dot is the local one. The MMSE can be read-off as the $\text{argmax} \Phi_{B}(E)$. The \emph{algorithmic threshold} $R_{\text{VAMP}}$ is indicated by the appearance of an inflexion point, i.e., appearance of a local maximum  (black curve). The  {\it information-theoretic threshold } $R_{\text{IT}}$ is obtained as the rate when the two maxima are equal (we never observed that more than two maxima are present).  When there is a unique maximum, such as when $R<R_{\text{VAMP}}$, or if the \emph{global} maximum is the rightmost one (at ``large'' $E$ value), namely when $R>R_{\text{IT}}$ (purple curve), then VAMP is conjectured to be asymptotically optimal: in both these scenarios VAMP matches the MMSE performance, but if $R<R_{\text{VAMP}}$ the decoding error is ``low'', while if $R>R_{\text{IT}}$ VAMP is still optimal but even the MMSE is poor. The presence of a local maximum whenever $R \in (R_{\text{VAMP}},R_{\text{IT}})$ (blue curve) prevents VAMP to decode:  a statistical-to-computational gap is present.}
    \label{fig:fig1}
\end{figure}

Fig. 1 shows that SE properly tracks VAMP for non-standard rotational invariant coding matrices. Under VAMP decoding, SS codes exhibit, as $L\!\to\!\infty$, a sharp phase transition at an \emph{algorithmic threshold} $R_{\text{VAMP}}$ below Shannon's capacity. $R_{\text{VAMP}}$ is defined as the highest rate such that for $R\!\le\!R_{\text{VAMP}}$, \eqref{SE} has a unique fixed point. For the gaussian ensemble $R^{\text{gauss}}_{\text{VAMP}} = 1.52$ and for the row-orthogonal one  $R^{\text{ortho}}_{\text{VAMP}} = 1.62$ and is therefore better compared with the gaussian coding matrices, as noted already for compressive sensing \cite{ma2014turbo}. Whenever $R<R_{\text{VAMP}}$ VAMP decodes well (and we conjecture optimally), see red and blue curves. If  instead $R\!>\!R_{\text{VAMP}}$ VAMP fails, see green and black curves.

Fig. \ref{fig:fig1} depicts the replica potential for the row-orthogonal coding ensemble  with $\text{snr} = 28$ and $B=2$ (a similar picture would appear for another ensemble). The advantage of the potential when compared to the SE analysis is that, in addition to encode $R_{\text{VAMP}}$ as the rate at which an inflexion point appears (which blocks the SE recursion seen as a gradient ascent of $\Phi_{B}(E)$ initialized at high $E$), it allows us to also obtain the \emph{information-theoretic threshold} of the code, defined as the rate where both the ``good'' and ``bad'' maxima of  $\Phi_{B}(E)$ are equal. If $R < R_{\text{VAMP}}$, VAMP's estimate is conjectured to match the MMSE estimator when initialized randomly, as in  standard gaussian SS codes or in linear regression \cite{barbier2017approximate,rush2017capacity,barbier2019optimal,reeves2016replica}. Instead, if $R > R_{\text{VAMP}}$ VAMP is  sub-optimal. 

\begin{figure}
    \centering
    \includegraphics[trim={104 40 98 70},clip,scale=0.42]{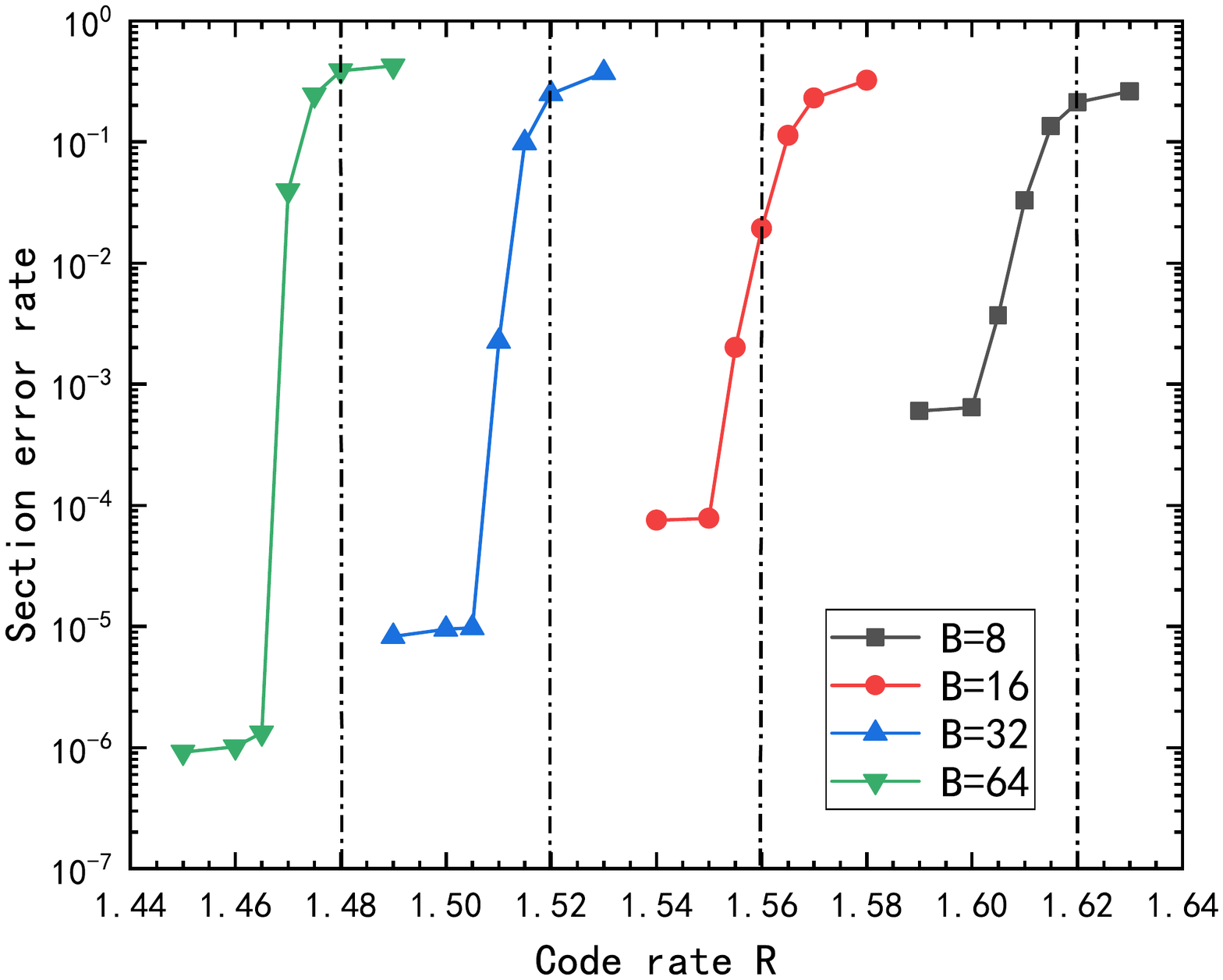}
    \caption{VAMP's performance quantified by its section error rate near its \emph{algorithmic threshold} (black dotted line) for $L=2^{16}, \text{snr}=15.0$ and various $B$. We use discrete cosine transform matrices as proxy of the row-orthogonal ensemble. The curves are averaged over $300$ realizations.}
    \label{fig:BP_threshold}
\end{figure}

At finite size VAMP's performance close its threshold $R_{\text{VAMP}}$, itself extracted from the potential as explained in the caption of Fig.~\ref{fig:fig1}, is shown in Fig.~\ref{fig:BP_threshold}. This was done using a proxy of the row-orthogonal ensemble based on  discrete cosine transform matrices. Using these structured matrices dramatically speeds-up the decoding under VAMP while having same performance, which is practically interesting. As predicted by our theory, recovery is good whenever $R<R_{\text{VAMP}}$ but poor else. When $B$ increases, the SER changes more drastically as the rate increases. Near its threshold and due to finite size effects, the VAMP performance (averaged over many realizations) has a transient behavior smoothly interpolating between very ``poor'' and very ``good''.

All  codes are made available at \cite{code}. Equipped with these methods, we can therefore completely characterize the performance of SS codes and VAMP with generic rotational invariant coding ensembles as $L\to\infty$.

\subsection{Analysis in the large section size limit, and main result}

The analysis in the large section size limit $B\to\infty$ requires rescaling the potential $\tilde{\Phi}(E) \defeq \lim_{B \to \infty} \Phi_{B}(E)/\ln B $ for it to possess a finite limit. The ``entropic contribution'' $\lim_{B \to \infty}  S_{B}(\Sigma)/\ln B = \max (1 - \tilde\Sigma^{-2}/2, 0 )$ has been computed in \cite{barbier2017approximate} using the replica method, where the effective SNR that needs to be rescaled too is, using $R=\log_2(B)/(\alpha B)$, given by  $\tilde \Sigma (E)^{-2} := \lim_{\alpha \to 0} \text{snr} \mathcal{R}(- \text{snr} E)/( \alpha R \ln 2)$; note that for $R$ to remain finite as $B\to\infty$ then necessarily $\alpha =\Theta(\ln B /B)\to 0$ and that moreover also $\mathcal{R}$ depends on $\alpha$. As $\alpha \to 0$ we Taylor expand $\mathcal{C}^{-1}(-z) = z^{-1} + \Psi(z)  \alpha + o(\alpha) $ and thus $\mathcal{R}(z)/\alpha=\Psi(z)+o_\alpha(1)$. 
%
So $\tilde{\Phi}(E)$ has a well defined expression for small $\alpha$ (i.e., large $B$) given by
\begin{equation*}
    \max \Big(1,\frac{1}{2 \tilde \Sigma(E)^2}\Big) - \frac{1-E}{ 2\tilde\Sigma(E)^2} - \int_0^{\text{snr} E} \frac{ \Psi(-u) }{ 2 R \ln 2}\mathrm{d} u+o_\alpha(1).
\end{equation*}
The expressions of $\Psi$ for the three spectra we focus on (in order) are worked out easily and read as follows:
 $\Psi_{\text{gauss}} (z) =\Psi_{\text{ortho}}(z) = (1-z)^{-1} $, while $ \Psi_{\text{discrete}}(z)=(4-3z)/((z-2)(3z-2))$. 
 From the analysis of $\tilde{\Phi}$ we can extract our main result stated below (derived in the next section). Notice that $\text{rank}(\textbf{A})=\alpha N$ with $\alpha\le 1$, so we can rewrite the asymptotic spectral density of $B^{-1} \textbf{A}^\intercal \textbf{A}$ as $\rho = (1-\alpha) \delta_0 + \alpha \rho_{\text{supp}}$
where $\rho_{\text{supp}}$ is a p.d.f. of mean $1$ due to the power constraint.
\newtheorem{theorem}{Result}
\begin{theorem}  \label{th:th1}
Consider SS codes with coding matrix $\textbf{A}$ drawn from a rotational invariant ensemble, whose empirical spectral measure converges to a well defined density with finite support as $L\to \infty$. Then the $B \to +\infty$ limit $R_{\text{VAMP}} (\infty)$ of the VAMP threshold verifies $R_{\text{VAMP}} (\infty) \leq \text{snr}/(2 (1+\text{snr}) \ln 2)$. Moreover the code is capacity achieving in the sense that the infinite section size limit $R_{\text{IT}} (\infty)$ of the information-theoretic threshold satisfies      $R_{\text{IT}} (\infty) = \log_{2} (1+\text{snr})/2 := C$, with $ C $ the Shannon capacity of the AWGNC, if and only if the asymptotic p.d.f. $\rho_{\text{supp}}$ of the non-zero eigenvalues of $B^{-1} \textbf{A}^\intercal \textbf{A}$ verifies $\rho_{\text{supp}} \to \delta_1$ in law when $B\to\infty$, $\alpha \to 0$. Moreover, in that case the algorithmic threshold is as good as it can be, i.e., $R_{\text{VAMP}} (\infty) = \text{snr}/(2 (1+\text{snr}) \ln 2)$.
\end{theorem}

\begin{figure}
    \centering
    \includegraphics[trim={110 40 98 70},clip,scale=0.43]{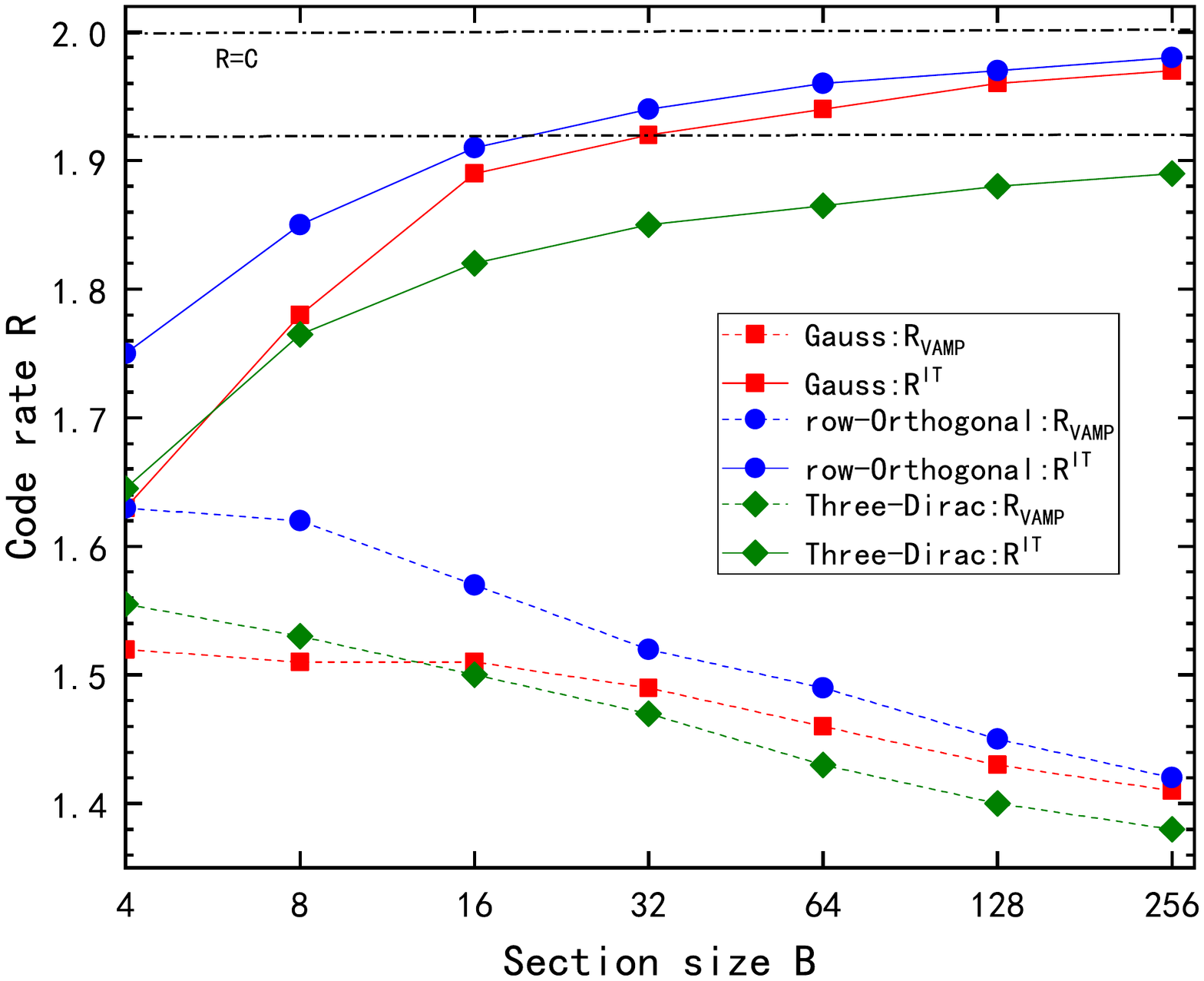}
    \caption{Algorithmic $R_{\text{VAMP}}$ and information-theoretic $R_{\text{IT}}$ thresholds as a function of the section size $B$ when $\text{snr} = 15$. The top dashed line is the channel capacity $C$ and the second top dashed line represents $R^{\text{discrete}}_{\text{IT}}(B=\infty)$. The information-theoretic threshold for the row-orthogonal ensembles approaches $C$ faster than the gaussian ensemble which is interesting in practice (despite that when $B\to+\infty$ they both converge to $C)$, while the discrete spectrum ensemble does not saturate Shannon's limit, even as $B\to\infty$ (in which case $R_{\text{VAMP}}^{\text{gauss}}(\infty) = R_{\text{VAMP}}^{\text{ortho}}(\infty)\approx 0.67$ and  $R_{\text{VAMP}}^{\text{discrete}}(\infty) \approx 0.61$). Thus, in the large section size limit, row-orthogonal matrices are not better than Gaussian ones; already at $B=256$ their thresholds are very similar. However, there may be other benefits to using row-orthogonal matrices, e.g., rate of convergence of VAMP decoding and convergence to lower error (see Fig. \ref{fig:SE of VAMP}). }
    \label{fig:fig2}
\end{figure}

According to this ``spectral criterion'' of Result \ref{th:th1}, both the gaussian and row-orthogonal coding ensembles are capacity-achieving in the large section size limit (taken after the $L\to\infty$ limit), while the discrete spectrum is not. E.g., when $\text{snr} = 15$ we can extract from potential $\tilde{\Phi}$ (see the details in the derivation of Result \ref{th:th1}) that  $R^{\text{gauss}}_{\text{IT}}(\infty)=R^{\text{ortho }}_{\text{IT}}(\infty) = C= 2$ while $R^{\text{discrete}}_{\text{IT}}(\infty) = 1.91$. The algorithmic and information-theoretic thresholds extracted at finite section size $B$ (but infinite $L$) are shown in Fig.~\ref{fig:fig2}, do converge when $B$ increases to their predicted asymptotics.
This criterion strongly suggests that the row-orthogonal ensemble is optimal among rotationally invariant ensembles for coding in SS codes, at least information-theoretically, given that $\rho_{\text{supp}}=\delta_1$ even for finite $B$. Fig.~\ref{fig:fig2} also indicates that also the VAMP threshold seems better than with other ensembles.

\subsection{Derivation of Result \ref{th:th1} by a replica analysis}

The derivation of Result \ref{th:th1} relies on an auxiliary lemma:
\newtheorem{lemma}{Lemma}
\begin{lemma} \label{le:lemma1}
   For $z\in \mathbb{R}_{< 0}$: $(i)$  $\mathcal{R}^{\prime} (z) > 0$; $(ii)$  $\Psi(z) \leq \frac{1}{1 - z}.$
\end{lemma}
\begin{proof}
    We start with $(i)$. $\mathcal{C}^{\prime} (z) = \int  d \lambda \frac{ \rho(\lambda)}{(\lambda - z)^2}   $ so $\mathcal{C}$ is strictly increasing. Therefore, the inverse $\mathcal{C}^{-1}(z): \mathbb{R}_{>0} \mapsto \mathbb{R}_{<0}$ is well defined. We have
   \begin{equation*}
       \mathcal{R}^{\prime} (z) = \frac{1}{z^2} - \frac{1}{\mathcal{C}^{\prime}(\mathcal{C}^{-1}(-z))}, \quad z<0.
   \end{equation*}
    For any $z<0$, let $t>0$ s.t. $z=- \mathcal{C}(t)$ ($t$ exists by monotony of $\mathcal{C}$). Then
   \begin{align*}
\mathcal{R}^{\prime} (- \mathcal{C}(t))=\frac{\mathcal{C}^{\prime}(t) - \mathcal{C}^2(t)}{\mathcal{C}^{\prime}(t) \mathcal{C}^2(t)} = \frac{\text{Var} ((\lambda - t)^{-1})}{\mathcal{C}^{\prime}(t) \mathcal{C}^2(t)}  >0.
   \end{align*}
     
We have proved $(i)$, and now consider $(ii)$. Recall that $\rho = (1-\alpha) \delta_0 + \alpha \rho_{\text{supp}}$, where $\rho_{\text{supp}}$ is the asymptotic law of the positive eigenvalues of $B^{-1}\textbf{A}^\intercal \textbf{A}$, and that the power constraint requires $ \int \lambda \rho_{\text{supp}} (\lambda) \mathrm{d} \lambda = 1.$ From $\mathcal{C}$'s definition we have 
\begin{equation}\label{2}
    -z = \frac{\alpha-1}{\mathcal{C}^{-1}(-z)} + \alpha\int \frac{\rho_{\text{supp} }(\lambda)}{\lambda - \mathcal{C}^{-1}(-z)} \mathrm{d} \lambda.
\end{equation}
Let $\rho_0$ (whose domain is $\mathbb{R}_{> 0})$ be the $\alpha \to 0$ limit of $\rho_{\text{supp}}$, and $0<\lambda_0\sim \rho_0$. Note that $\rho_0$ also satisfies the power constraint $ \mathbb{E}\lambda_0 =1$. Recall $\mathcal{C}^{-1}(-z)=z^{-1}+\Psi(z) \alpha+o(\alpha)$, so by multiplying both sides of \eqref{2} by $\mathcal{C}^{-1}(-z)/\alpha$ followed by letting $\alpha\to 0$ yields (we exchange limit and integration by dominated convergence)
\begin{equation} \label{eq:az}
    \Psi(z) z = \mathbb{E}\Big(\frac{1}{1-z\lambda_0}\Big) -1.
\end{equation}
By Cauchy-Schwarz we have
\begin{align*}  
    &1\le \mathbb{E}\Big(\frac{1}{1-z\lambda_0}\Big) \mathbb{E} (1- z\lambda_0 ) =\mathbb{E}\Big(\frac{1-z}{1-z\lambda_0}\Big)  ,
\end{align*}
where the equality holds if and only if $\rho_0 = \delta_{1}$. Combining this inequality with (\ref{eq:az}) proves $\Psi(z)\le \frac1{1-z}$ for $z<0$, with equality if and only if $\rho_{\text{supp}}  \to \delta_1$ as $\alpha \to 0$.
\end{proof}

We are in position to derive our main Result \ref{th:th1}.
\begin{proof}[Derivation of Result 1]
We consider various scenarios for the extrema of potential $ \tilde{\Phi}(E) $ in order to locate the two thresholds of interest (recall the caption of Fig. \ref{fig:fig1} for locating the transitions from ${\Phi}_B$, or from $\tilde{\Phi}$ at infinite $B$). Very similar analyses were performed in  \cite{barbier2017approximate,barbier2016threshold} so we will be brief. Start by noticing that $\tilde \Sigma (E)^{-2} := \lim_{\alpha \to 0} \text{snr} \mathcal{R}(- \text{snr} E)/( \alpha R \ln 2)$ is a decreasing function from Lemma~\ref{le:lemma1}, so $E = 1$ is its minimum (as $E\in[0,1]$).
\begin{itemize}
\item Case 1: $(2 \tilde\Sigma(1)^2)^{-1} > 1 $. Recall $\mathcal{R}(z)=\alpha \Psi(z)+o(\alpha)$. We have (and using Lemma 1 for the inequality)
\begin{equation} \label{eq:derivative 1}
\tilde{\Phi}^{\prime} (E) = -\lim_{\alpha \to 0} \frac{\text{snr}^2E}{2 \alpha R \ln 2}  \mathcal{R}^{\prime}(-\operatorname{snr} E) <0.
\end{equation}
There is a stable unique maximum at $E=0$.
\item Case 2: $(2 \tilde\Sigma(1)^2)^{-1} \le 1 $. There exists $E_1 \in [0,1]$ s.t. $(2 \tilde\Sigma(E_1)^2)^{-1} = 1$. The derivative of $\tilde{\Phi}(E)$ is (\ref{eq:derivative 1}) if $0<E<E_1$. When $E_1<E<1$ it is instead
\begin{equation} \label{eq:derivative 2}
    \tilde{\Phi}^{\prime}(E) = \lim_{\alpha \to 0}  \frac{\text{snr}^2}{2 \alpha R \ln 2} (1-E) \mathcal{R}^{\prime}(-\operatorname{snr} E) > 0.
\end{equation}
There are thus two maxima at $E = 0$ and $E=1$.
\end{itemize}
Solving $2 \tilde\Sigma(1)^2 = 1 $ for $R$ thus gives the algorithmic threshold in the large section limit:
\begin{equation} \label{eq:BP}
    R_{\text{VAMP}} (\infty) = \lim_{\alpha \to 0} \frac{\text{snr} \mathcal{R} (-\text{snr})}{2 \alpha \ln 2} = \frac{\text{snr} \Psi(-\text{snr})}{ 2 \ln 2}.
\end{equation}
Under case 2, the free entropy takes the following values at its maxima:
\begin{align*}
    \tilde{\Phi} (0) = 0, \ \ \tilde{\Phi} (1) = 1 - \lim_{\alpha \to 0} \frac{1}{2\alpha R \ln 2} \int_{0}^{{ {{\mathrm{snr}}} }}  \mathcal{R}(-u) \mathrm{d} u,
\end{align*}
Then setting $\tilde{\Phi}(0) = \tilde{\Phi} (1)$ gives the information-theoretic threshold in the large section limit:
\begin{equation}  \label{eq:OP}
    R_{\text{IT}}(\infty) = \lim_{\alpha \to 0} \frac{\int_{0}^{{{\mathrm{snr}}} } \mathcal{R}(-u) \mathrm{d} u }{2 \alpha \ln 2} = \frac{\int_{0}^{{{  \mathrm{snr} }} } \Psi(-u) \mathrm{d} u}{2 \ln 2} .
\end{equation}
With Lemma 1 and (\ref{eq:BP}), (\ref{eq:OP}), we can easily derive an spectrum-independent upper bound on both thresholds:
\begin{align*}
    &R_{\text{VAMP}}(\infty) \leq \frac{\text{snr}}{2(1+\text{snr}) \ln 2 },\\
    &R_{\text{IT}}(\infty) \leq \frac{1}{2} \log_{2} (1+\text{snr}) = C. 
\end{align*}
Both equalities hold if and only if $\rho_{\text{supp}}  \to \delta_1$ as $\alpha \to 0$ as claimed. In this case the scheme is capacity-achieving \emph{and} VAMP's algorithmic threshold is as good as it can be.
\end{proof}

\section{ Perspectives}

There are a number of natural extensions of the present work, in the spirit of recent developments in (generalized) regression with design matrices beyond i.i.d. gaussian \cite{barbier2018mutual,ma2021analysis,maillard2020construction,aubin2020exact}. For practical purposes, further studies should concentrate on the influence of the spectra of coding matrices for finite section size SS codes (while our analysis mainly focused on large $B$). It is also  interesting to investigate whether a similar criterion as Result 1 may be extended to SS codes for more generic memoryless channels \cite{barbier2016threshold,biyik2017generalized}. Another natural direction to explore concerns the comparison of the ``spectral design'' we proposed with  different types of structures for the coding matrices, in particular power allocation and spatial-coupling \cite{barbier2017approximate,rush2021capacity}. Or to analyze what happens when these structures are combined, e.g., when the blocks of the spatially-coupled matrices are themselves drawn from a rotational invariant ensemble \cite{wen2014analysis}, and see whether threshold saturation \cite{barbier2016threshold,barbier2016proof} (i.e., the ``closing'' of the computational-statistical gap of AMP-based decoders) occurs in that setting.   

\section{Acknowledgements}
We thank Galen Reeves, Shansuo Liang, Hao Wu and Zhongyi Huang for helpful discussions.



\begin{thebibliography}{10}

\bibitem{barron2010toward}
A.R. Barron and A.~Joseph.
\newblock Toward fast reliable communication at rates near capacity with
  gaussian noise.
\newblock In {\em 2010 IEEE International Symposium on Information Theory},
  pages 315--319, 2010.

\bibitem{rangan2019vector}
S.~Rangan, P.~Schniter, and A.K. Fletcher.
\newblock Vector approximate message passing.
\newblock {\em IEEE Trans. on Information Theory}, 65(10):6664--6684, 2019.

\bibitem{rush2017capacity}
C.~Rush, A.~Greig, and R.~Venkataramanan.
\newblock Capacity-achieving sparse superposition codes via approximate message
  passing decoding.
\newblock {\em IEEE Trans. on Information Theory}, 63(3):1476--1500, 2017.

\bibitem{barbier2017approximate}
J.~Barbier and F.~Krzakala.
\newblock Approximate message-passing decoder and capacity achieving sparse
  superposition codes.
\newblock {\em IEEE Transactions on Information Theory}, 63(8):4894--4927,
  2017.

\bibitem{barbier2016proof}
J.~Barbier, M.~Dia, and N.~Macris.
\newblock Proof of threshold saturation for spatially coupled sparse
  superposition codes.
\newblock In {\em 2016 IEEE International Symposium on Info. Theory (ISIT)},
  pages 1173--1177, 2016.

\bibitem{biyik2017generalized}
E.~Biyik, J.~Barbier, and M.~Dia.
\newblock Generalized approximate message-passing decoder for universal sparse
  superposition codes.
\newblock In {\em 2017 IEEE International Symposium on Information Theory
  (ISIT)}, pages 1593--1597, 2017.

\bibitem{rush2021capacity}
C.~Rush, K.~Hsieh, and R.~Venkataramanan.
\newblock Capacity-achieving spatially coupled sparse superposition codes with
  amp decoding.
\newblock {\em IEEE Transactions on Information Theory}, 2021.

\bibitem{ma2017orthogonal}
J.J Ma and L.~Ping.
\newblock Orthogonal amp.
\newblock {\em IEEE Access}, 5:2020--2033, 2017.

\bibitem{mezard2009information}
M.~M\'ezard and A.~Montanari.
\newblock {\em Information, physics, and computation}.
\newblock Oxford University Press, 2009.

\bibitem{bayati2011dynamics}
M.~Bayati and A.~Montanari.
\newblock The dynamics of message passing on dense graphs, with applications to
  compressed sensing.
\newblock {\em IEEE Transactions on Information Theory}, 57(2):764--785, 2011.

\bibitem{javanmard2013state}
A.~Javanmard and A.~Montanari.
\newblock State evolution for general approximate message passing algorithms,
  with applications to spatial coupling.
\newblock {\em Information and Inference: A Journal of the IMA}, 2(2):115--144,
  2013.

\bibitem{barbier2020mutual}
J.~Barbier, N.~Macris, M.~Dia, and F.~Krzakala.
\newblock Mutual information and optimality of approximate message-passing in
  random linear estimation.
\newblock {\em IEEE Trans. on Information Theory}, 66(7):4270--4303, 2020.

\bibitem{reeves2016replica}
G.~Reeves and H.D. Pfister.
\newblock The replica-symmetric prediction for compressed sensing with gaussian
  matrices is exact.
\newblock In {\em 2016 IEEE International Symposium on Info. Theory (ISIT)},
  pages 665--669, 2016.

\bibitem{barbier2016threshold}
J.~Barbier, M.~Dia, and N.~Macris.
\newblock Threshold saturation of spatially coupled sparse superposition codes
  for all memoryless channels.
\newblock In {\em 2016 IEEE Info. Theory Workshop (ITW)}, pages 76--80, 2016.

\bibitem{barbier2019optimal}
J.~Barbier, F.~Krzakala, N.~Macris, L.~Miolane, and L.~Zdeborov{\'a}.
\newblock Optimal errors and phase transitions in high-dimensional generalized
  linear models.
\newblock {\em Proceedings of the National Academy of Sciences},
  116(12):5451--5460, 2019.

\bibitem{barbier2018mutual}
J.~Barbier, N.~Macris, A.~Maillard, and F.~Krzakala.
\newblock The mutual information in random linear estimation beyond iid
  matrices.
\newblock In {\em 2018 IEEE International Symposium on Information Theory
  (ISIT)}, pages 1390--1394, 2018.

\bibitem{berthier2020state}
R.~Berthier, A.~Montanari, and P.-M. Nguyen.
\newblock State evolution for approximate message passing with non-separable
  functions.
\newblock {\em Information and Inference: A Journal of the IMA}, 9(1):33--79,
  2020.

\bibitem{tulino2004random}
A.~Tulino and S.~Verd{\'u}.
\newblock {\em Random matrix theory and wireless communications}.
\newblock Now Publishers Inc, 2004.

\bibitem{potters2020first}
Marc Potters and Jean-Philippe Bouchaud.
\newblock {\em A First Course in Random Matrix Theory: For Physicists,
  Engineers and Data Scientists}.
\newblock Cambridge University Press, 2020.

\bibitem{tulino2013support}
A.M. Tulino, G.~Caire, S.~Verd\'u, and S.~Shamai.
\newblock Support recovery with sparsely sampled free random matrices.
\newblock {\em IEEE Transactions on Information Theory}, 59(7):4243--4271,
  2013.

\bibitem{gerbelot2020asymptotic}
C.~Gerbelot, A.~Abbara, and F.~Krzakala.
\newblock Asymptotic errors for teacher-student convex generalized linear
  models (or: How to prove kabashima's replica formula).
\newblock {\em arXiv preprint arXiv:2006.06581}, 2020.

\bibitem{fan2022approximate}
Z.~Fan.
\newblock Approximate message passing algorithms for rotationally invariant
  matrices.
\newblock {\em The Annals of Statistics}, 50(1):197--224, 2022.

\bibitem{dudeja2020information}
R.~Dudeja, J.~Ma, and A.~Maleki.
\newblock Information theoretic limits for phase retrieval with subsampled haar
  sensing matrices.
\newblock {\em IEEE Transactions on Information Theory}, 66(12):8002--8045,
  2020.

\bibitem{venkataramanan2021estimation}
R.~Venkataramanan, K.~K{\"o}gler, and M.~Mondelli.
\newblock Estimation in rotationally invariant generalized linear models via
  approximate message passing.
\newblock {\em arXiv preprint arXiv:2112.04330}, 2021.

\bibitem{barbier2014replica}
J.~Barbier and F.~Krzakala.
\newblock Replica analysis and approximate message passing decoder for
  superposition codes.
\newblock In {\em 2014 IEEE International Symposium on Information Theory},
  pages 1494--1498, 2014.

\bibitem{ma2014turbo}
J.J. Ma, X.J. Yuan, and L.~Ping.
\newblock Turbo compressed sensing with partial dft sensing matrix.
\newblock {\em IEEE Signal Processing Letters}, 22(2):158--161, 2014.

\bibitem{code}
T.~Hou, Y.~Liu, T.~Fu, and J.~Barbier.
\newblock
  \url{https://github.com/yztfu/VAMP-decoder-and-capacity-achieving-SS-code}.

\bibitem{ma2021analysis}
J.J. Ma, J.~Xu, and A.~Maleki.
\newblock Analysis of sensing spectral for signal recovery under a generalized
  linear model.
\newblock {\em Advances in Neural Information Processing Systems}, 34, 2021.

\bibitem{maillard2020construction}
A.~Maillard, F.~Krzakala, Y.M. Lu, and L.~Zdeborov{\'a}.
\newblock Construction of optimal spectral methods in phase retrieval.
\newblock {\em arXiv preprint arXiv:2012.04524}, 2020.

\bibitem{aubin2020exact}
B.~Aubin, B.~Loureiro, A.~Baker, F.~Krzakala, and L.~Zdeborov{\'a}.
\newblock Exact asymptotics for phase retrieval and compressed sensing with
  random generative priors.
\newblock In {\em Mathematical and Scientific Machine Learning}, pages 55--73.
  PMLR, 2020.

\bibitem{wen2014analysis}
C.K. Wen and K.K. Wong.
\newblock Analysis of compressed sensing with spatially-coupled orthogonal
  matrices.
\newblock {\em arXiv:1402.3215}, 2014.

\end{thebibliography}

\end{document}